\documentclass[aps,pre,showpacs,twocolumn, showemail]{revtex4}
\usepackage{bbm}
\usepackage{mathrsfs}
\usepackage{epsfig,psfrag}
\usepackage{amsmath,amsfonts,amssymb}
\usepackage[usenames]{color}
\usepackage[dvips, bookmarks, colorlinks=true, plainpages = false, citecolor = blue, linkcolor = blue, urlcolor = blue, filecolor = blue]{hyperref}

\def\be{\begin{equation}}
\def\ee{\end{equation}}
\def\bea{\begin{eqnarray}}
\def\eea{\end{eqnarray}}

\begin{document}

\preprint{draft}

\title{Scaling of Clusters and Winding Angle Statistics of Iso-height Lines in two-dimensional KPZ Surface}
\author{  A. A. Saberi $^1$}\email{a$_$saberi@ipm.ir} \author{S. Rouhani  $^2$}
\address
{$^1$ School of Physics, Institute for Studies in Theoretical
Physics and Mathematics, 19395-5531 Tehran, Iran \\$^2$ Department
of Physics, Sharif University of Technology, P.O. Box 11155-9161,
Tehran, Iran }
\date{\today}

\begin{abstract}
We investigate the statistics of Iso-height lines of
(2+1)-dimensional Kardar-Parisi-Zhang model at different level sets
around the mean height in the saturation regime. We find that the
exponent describing the distribution of the height-cluster size
behaves differently for level cuts above and below the mean height,
while the fractal dimensions of the height-clusters and their
perimeters remain unchanged. The winding angle statistics also
confirms again the conformal invariance of these contour lines in
the same universality class of self-avoiding random walks (SAWs).
\end{abstract}


\maketitle

The study of surfaces, their static statistical properties, as well
as growth and evolution dynamics and their morphological properties,
has been attracting an ever increasing amount of interest over the
last two decades.\\ A method of characterizing surfaces is by
looking at their iso-height contour lines and the islands that are
generated by a cut through the surface at a certain constant height.
Such a situation can be found in the patterns exhibited by
topographical islands and continents. These islands and their
coastlines have some fractal features with a fractal dimension
related to the roughness exponent $\alpha$, given by the structure
function $\langle[h(x)-h(x+r)]^2\rangle\sim r^{2\alpha}$, where
$h(x)$ is the height of interface as a function of its position.

Theoretical modeling of the growth processes started with the work
by Edwards and Wilkinson (EW) \cite{EW} who suggested that one might
describe the dynamics of the height fluctuations by a simple linear
diffusion equation. Kardar, Parisi, and Zhang (KPZ) \cite{Kardar}
realized that there is a relevant term proportional to the square of
the height gradient which represents a correction for lateral
growth. The KPZ equation is given by \be\label{eqKPZ} \frac{\partial
h(\textbf{x},t)}{\partial t}=\nu\nabla^2h+\frac{\lambda}{2}
\mid\nabla h\mid^2+\eta(\textbf{x},t)\;. \ee The first term on the
r.h.s describes relaxation of the interface caused by a surface
tension $\nu$, and the nonlinear term is due to the lateral growth.
The noise $\eta$ is uncorrelated Gaussian white noise in both space
and time with zero average i.e., $
\langle\eta(\textbf{x},t)\rangle=0 $ and $\langle
\eta(\textbf{x},t)\eta(\textbf{x}',t')\rangle=2D\delta^d(\textbf{x}-\textbf{x}')\delta(t-t')$.

In (1+1)-dimension the roughness and growth exponents were exactly
obtained, $\alpha=1/2$ and $\beta=1/3$ respectively \cite{stanley},
while for the (2+1)-dimensional case there are just numerical
evidence and predictions \cite{Parisi}.

The KPZ equation is invariant under translations along both growth
direction and perpendicular to it, as well as time translation and
rotation. Despite these symmetries, the existence of driving force
perpendicular to the interface (due to the nonlinear term), breaks
the up-down symmetry ($h\rightarrow-h$) \cite{stanley}. In two
dimensions these symmetries and growth dynamics can affect the
statistics of the iso-height lines (island coastlines) at different
level sets, which is the main subject of the present work.

In our previous paper \cite{saberi} we have focused on zero
iso-height lines of the (2+1)-dimensional KPZ model in the
saturation regime (mean height was set to zero). Using the theory of
Schramm-Loewner evolution (SLE), we have numerically shown that the
contour lines of zero height behave statistically like self avoiding
walks (SAWs) and they can be defined by the family of conformally
invariant curves i.e., SLE$_\kappa$ curves with diffusivity
$\kappa=8/3$. The statistics of these objects for the EW model has
been shown to be the same as the interfaces in the $O(2)$ model, and
can be described by SLE$_4$.

The SLE process, introduced by Schramm \cite{schramm} describes the
scaling limit of a variety of statistical mechanical models in two
dimensions (some review articles are given in \cite{cardy}). Schramm
and Sheffield showed that the contour lines in a two-dimensional
discrete Gaussian free field are statistically equivalent to
$SLE_{4}$ \cite{Schramm-Sheffield}. Moreover, it is shown that the
restriction property only applies in the case for $\kappa=8/3$
\cite{Schramm-Lawler}. Since self-avoiding random walk (SAW)
satisfies the restriction property, it is conjectured that in the
scaling limit it falls in the SLE class with $\kappa=8/3$
\cite{Tom-Kennedy}. The scaling limit of SAW in the half-plane has
been proven to exist \cite{G. Lawler} but there is no general proof
of its existence.

The theory of SLE has recently been applied to many experimental and
physical systems. It is shown that the statistics of the
zero-vorticity lines in inverse cascade of two dimensional (2D)
Navier-Stokes turbulence is conformally invariant and belongs to the
percolation universality class \cite{bernard1}. The same issue has
been studied for zero-temperature isolines in the inverse cascade of
surface quasigeostrophic turbulence \cite{bernard2}, domain walls of
spin glasses \cite{spin glass} and the nodal lines of random wave
functions \cite{Keatin}. Moreover, it has been shown recently that
the statistics of the iso-height lines on the experimentally grown
WO$_3$ surface is the same as domain walls statistics in the
critical Ising model \cite{WO3}. Avalanche frontiers in sandpile
models have been shown to be conformally invariant and in the same
universality class of loop erased
random walks  \cite{ASM}.\\

\begin{figure}\begin{center}
\includegraphics[scale=0.43]{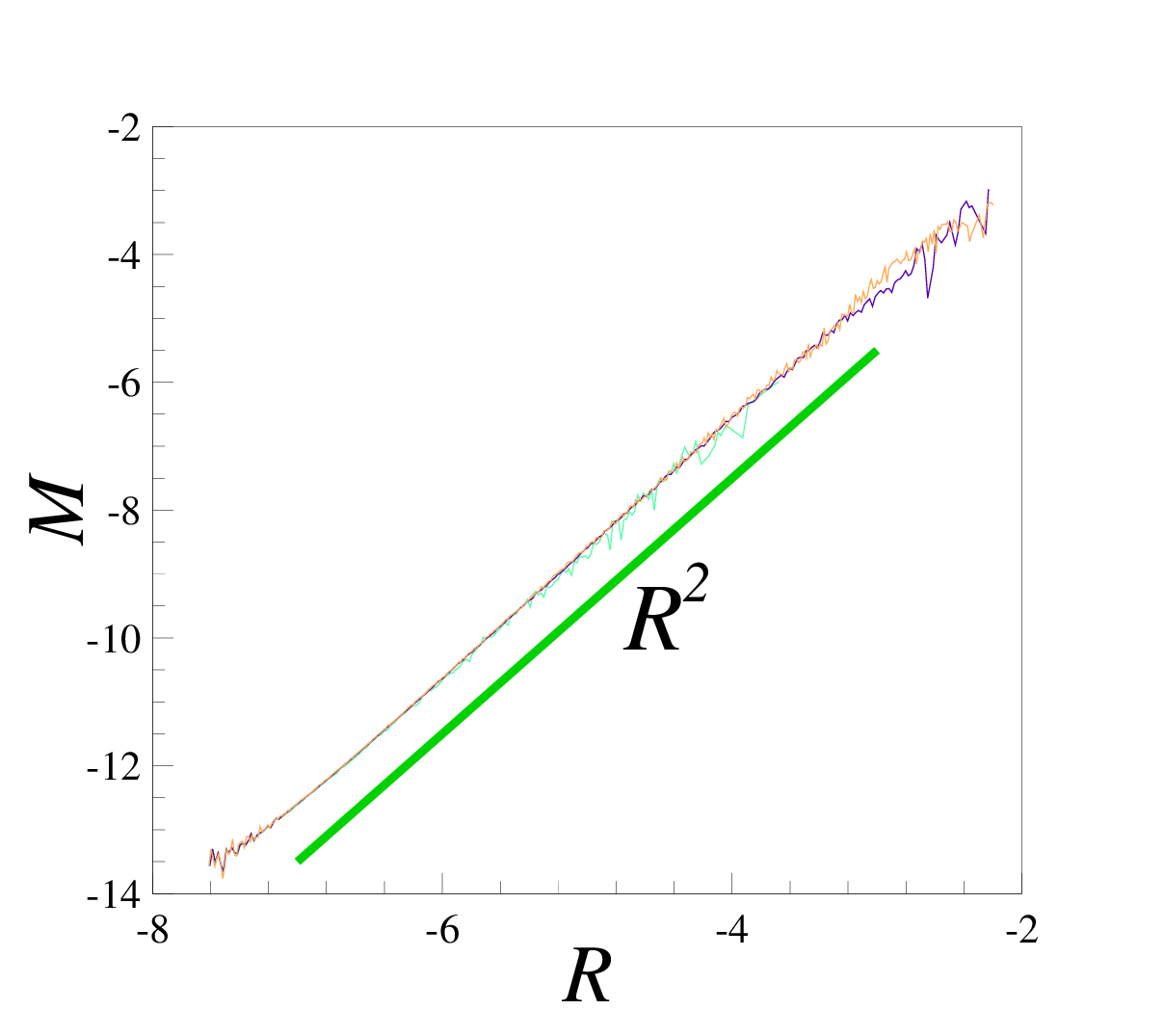}
\narrowtext\caption{\label{Fig1}(color online). Log-log plot of the
average area of a cluster $M$ versus the radius of gyration $R$, at
three different cuts with $\delta$ = -2, 0 and +2.}\end{center}
\end{figure}

\begin{figure}\begin{center}
\includegraphics[scale=0.43]{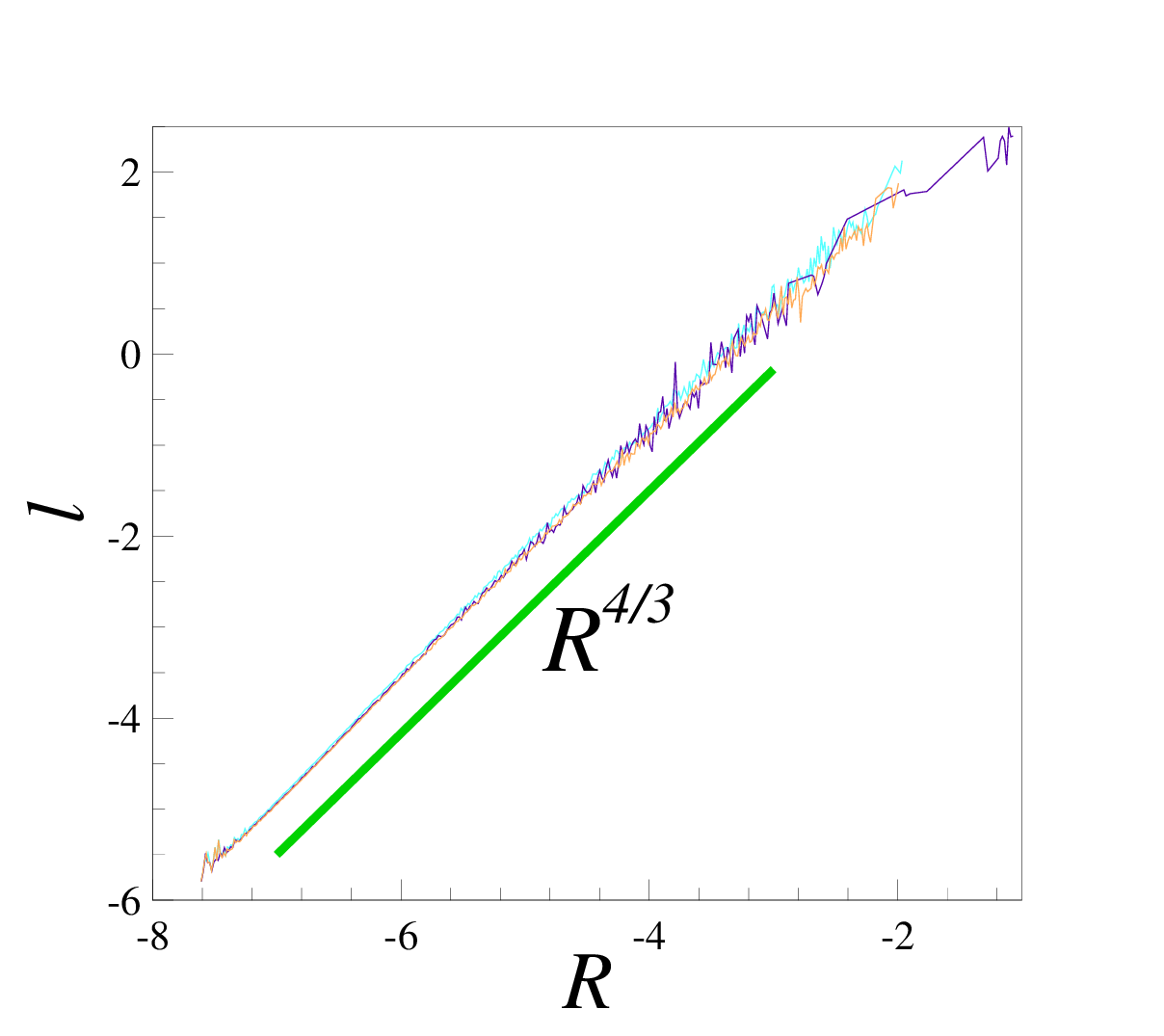}
\narrowtext\caption{\label{Fig2}(color online). Log-log plot of the
average length of a loop $l$ versus the radius of gyration $R$, at
three different cuts with $\delta$ = -2, 0 and +2.}\end{center}
\end{figure}

Here we briefly review the results obtained in \cite{saberi} for
zero height level cuts, and then extend them for cuts
made at different heights.\\
We have integrated the discretized KPZ equation on a square lattice
of size $2048^2$, with periodic boundary conditions. The details of
numerical integration and simulation are given in \cite{saberi}.\\
Consider an ensemble of 2D-KPZ saturated surfaces and a cut is made
at specific height say $h_\delta=\langle h(x)\rangle
+\delta\sqrt{\langle[h(x)-\langle h(x)\rangle]^2\rangle}:=0$, where
the symbol $\langle\cdot\cdot\rangle$ denotes spatial averaging.
Then we define each island (cluster height) as a set of connected
sites with positive height which were identified by the
Hoshen-Kopelman algorithm. \\The scaling of the mass $M$ of a
cluster with the radius of gyration $R$, behaves like $M\sim
R^{D_c}$, where $D_c$ is the fractal dimension of the cluster which
is $D_c=2$ in this case. As shown in Fig. \ref{Fig1}, this fractal
dimension remains unchanged
for different $\delta$.\\
The fractal dimension of a coastline (or loop because of periodic
boundary conditions), can be obtained with the scaling relation
between the average length of a loop $l$, and the radius of gyration
$R$, as $l\sim R^{D_l}$. This fractal dimension remains also fixed,
within numerical errors, for cuts made at different $\delta$. The
best fits to data shown in Fig. \ref{Fig2}, yields the fractal
dimension of a contour line in the range of $D_l=1.34\pm 0.02$ (Fig. \ref{Fig2}).\\
Powerful scaling arguments made by Kondev and Henley \cite{kondev},
connect the fractal dimension of a contour line to the roughness
exponent $\alpha$ of the surface, \be\label{kondev} D_l=2 - x_l -
\alpha/2,\ee where $x_l$ is the loop correlation exponent. Although
the exact value of $x_l=1/2$ is for $\alpha=0$ and $1$,
\cite{kondev1} but it is conjectured that its value is super
universal and is independent of $\alpha$ for Gaussian surfaces.\\
In the case of 2D-KPZ surface the finite size scaling for the
interface width yields the roughness exponent \cite{saberi}
$\alpha=0.37\pm0.01$, which is in mild conflict of Eq.
(\ref{kondev}). This may be because the field $h(x)$ does not follow
a Gaussian distribution. In other words, the fractal dimension
$D_l=1.34\pm 0.02\sim 4/3$ obtained for contour lines of 2D-KPZ
surface, is equal to what one obtains for Gaussian surfaces with
$\alpha=1/3$.

\begin{figure}\begin{center}
\includegraphics[scale=0.45]{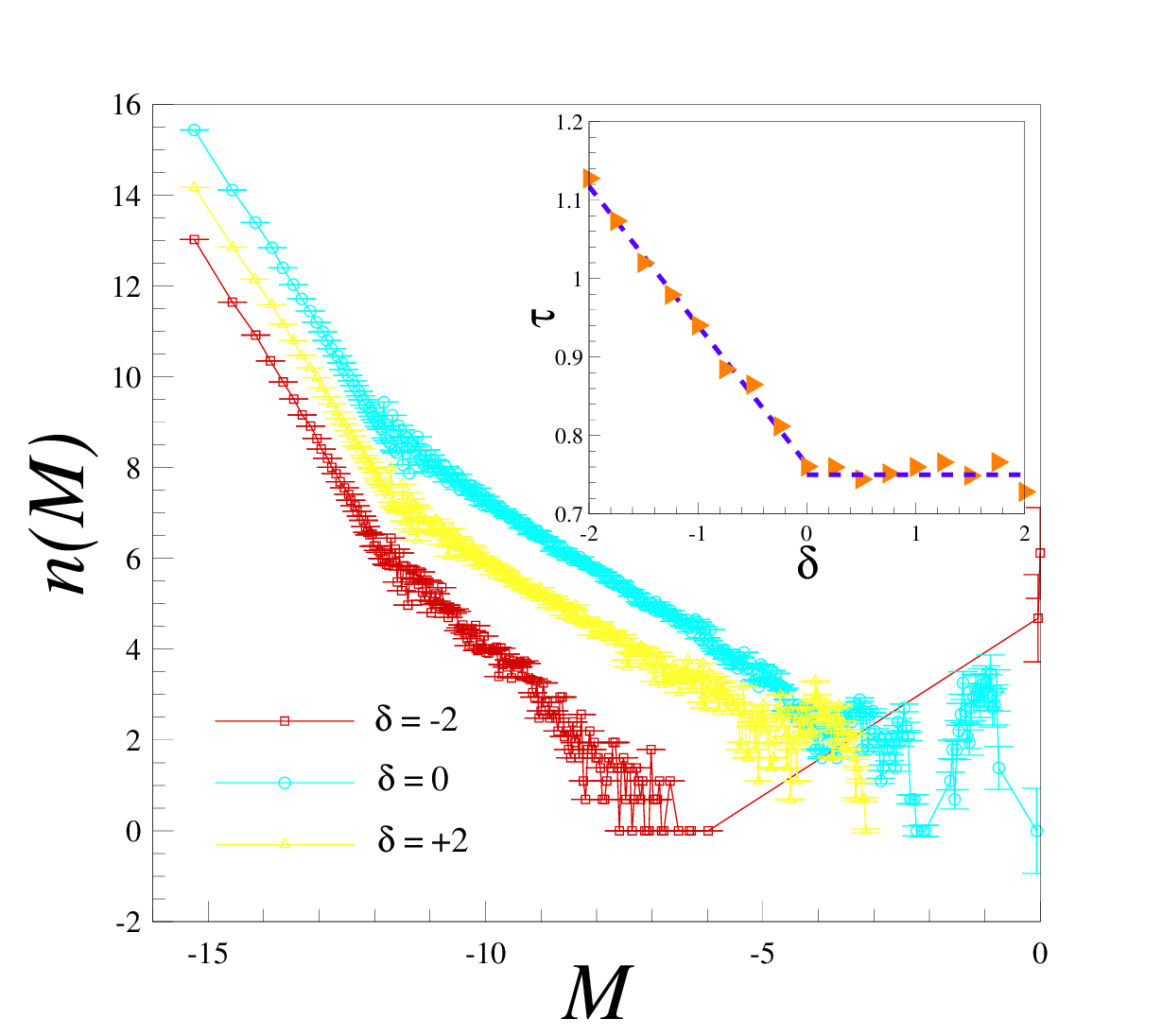}
\narrowtext\caption{\label{Fig3}(color online). Main: log-log plot
of the number of clusters of area between $M$ and $1.05 M$, at three
different cuts with $\delta$ = -2, 0 and +2. Inset: the exponents
for the island distributions as a function of $\delta$. The errors
are less than 0.05 for all exponents. The slope of the dashed lines
are $-1.77\pm 0.03$, the best fitted for $\delta<0$, and $0$, which
is drawn for comparison for $\delta\geq0$ at
$\tau=0.75$.}\end{center}
\end{figure}

The island size distribution has also a power-law behavior. As can
be seen in Fig. \ref{Fig3}, there are two distinct scaling regions
for the distribution of the island size. We find that the small size
islands are distributed according to a power-law distribution
$n(M)\sim M^{-\tau_s}$, with a same exponent $\tau_s=2\pm 0.05$ for
level cuts at different $\delta$.

For the other region with larger island size dominant, the power-law
behavior is held, but with different exponents at different level
cuts.\\ For cuts made at lower values of $\delta$, percolative
height clusters appear dominantly which their sizes are in the order
of sample size. Inside these percolative islands there are some
lakes (negative height clusters) which can also contain smaller
islands with the distribution as shown in Fig. \ref{Fig3} (for
$\delta=-2$). Increasing in the height of the cut, tends to a
continuous distribution of islands with different sizes and
different scaling behavior. We find that for cuts below the mean
height (i.e., $\delta<0$), the island-size distribution exponent
$\tau$ takes different values for different $\delta$, while it takes
almost a same value (within statistical error) for positive height
cuts. This behavior is shown in the inset of Fig. \ref{Fig3}, which
attributes two different regimes for level cuts i.e., $\delta<0$ and
$\delta\geq0$. For level cuts made at $\delta<0$, the island-size
distribution exponent $\tau$ decreases linearly with the slope of
$-1.77\pm 0.03$, while it takes values around $\tau=0.75$ within
statistical errors for $\delta\geq0$. This asymmetrical behavior of
the exponent $\tau$ around the mean height may be interpreted as the
breakdown of the up-down symmetry under changing $h\rightarrow-h$ in
the KPZ equation (\ref{eqKPZ}). This asymmetry in the dynamics of
the growth process can tend to an asymmetry in the statistics and
the distribution of valleys and overhangs.

\begin{figure}\begin{center}
\includegraphics[scale=0.47]{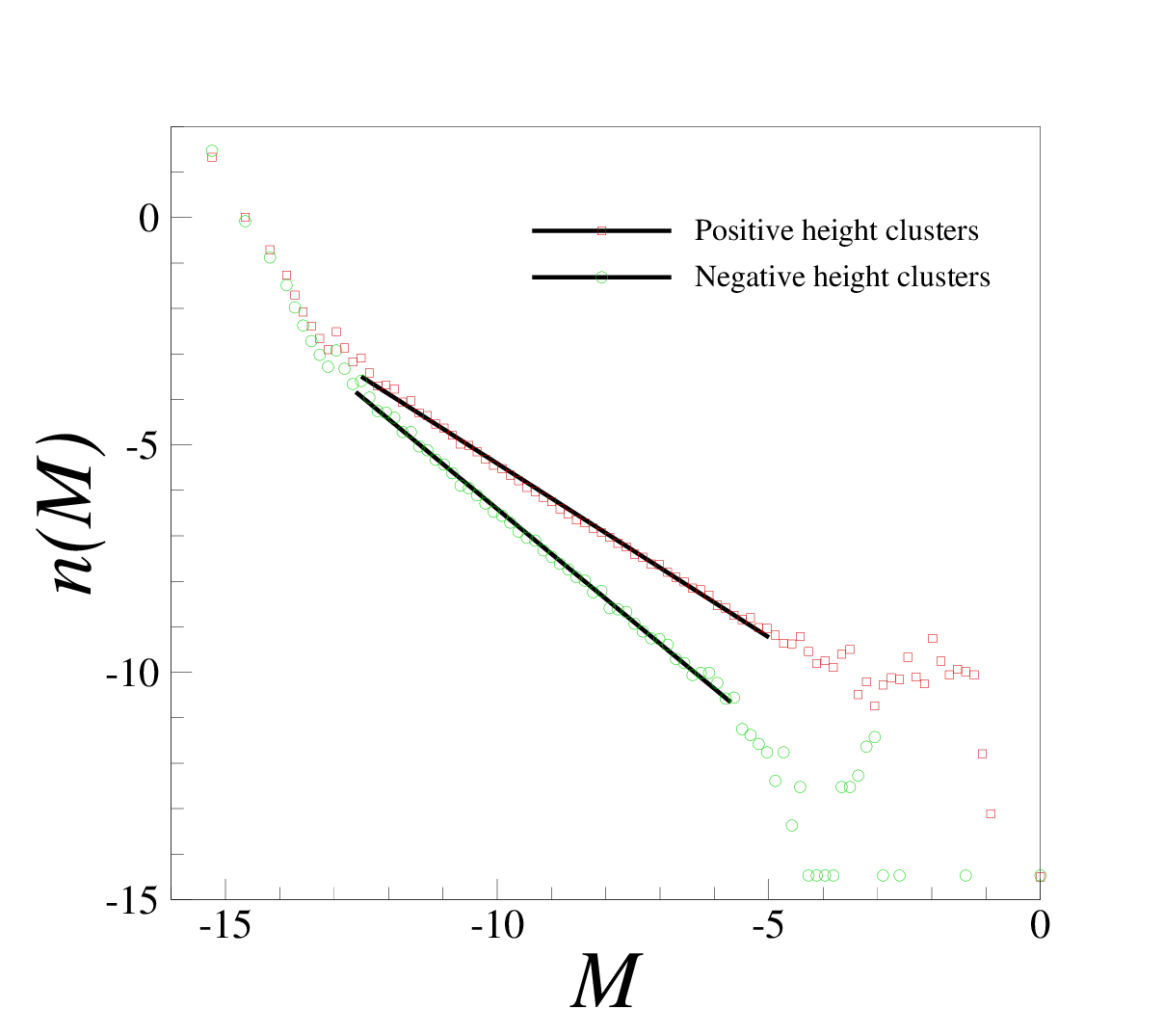}
\narrowtext\caption{\label{Fig4}(color online). Log-log plot of the
number of positive (squares) and negative (circles) height-clusters
of area between $M$ and $1.1 M$, at a level cut made at the mean
height with $\delta$ = 0. Solid lines are the best fits to the data
in the linear region yielding the distribution exponent
$\tau\sim0.77$ and $\tau\sim1$ for positive and negative
height-clusters, respectively.}\end{center}
\end{figure}

In a system with up-down symmetry it is expected that the valleys
and overhangs have the same statistical behavior with same
distribution. This behavior is not seen in the mean-height cuts of
the 2D-KPZ surface. The size distribution of the islands with
positive and negative heights differ. As shown in Fig. \ref{Fig4},
the exponents defining these two distributions are quite different
with values $\tau\sim0.77$ and $\tau\sim1$ for positive and negative
height-clusters, respectively.

Although the exact cause for the behavior of the exponent $\tau$ as
function of the height of the cut eludes us, but it may be possible
to guess its behavior around the mean height \cite{kondev} (a
slightly different problem has been investigated in \cite{Olami} for
random Gaussian surfaces). \\Using scaling arguments, it is shown in
\cite{kondev} that the average number density of contour lines (the
coastlines here), scales with the radius of gyration $R$ as
$n(R)\sim R^{-2+\alpha}$. Since the dimension of the islands in our
case is $2$ (see Fig. \ref{Fig1}), one can expect that the radius of
gyration for the perimeter of the islands and the islands themselves
have the same statistical behavior. So, the average number density
of the islands of size $M$, is given by $n(M)\sim M^{-1+\alpha/2}$.
Within an uncertainty in determining the roughness exponent
$\alpha$, it yields an approximated value for the exponent $\tau\sim
0.81$ which works here for the height cuts around and above the mean
height.

In the rest of the paper we investigate the conformal symmetry of
the coastlines of cuts made at different heights of two dimensional
saturated KPZ surface. Theory of SLE provides an appropriate
approach to check conformal properties of the geometrical features
of such systems. Looking at the fractal dimension obtained for the
KPZ coastlines at different level sets, it agrees with the SLE
curves of fractal dimension $D_f=1+\kappa/8$, with $\kappa=8/3$,
conjectured to describe the scaling limit of SAWs. In \cite{saberi},
we checked various consistencies between the coastlines and both
SAWs and SLE$_{8/3}$. Such coastlines can statistically be defined
as the outer boundary of the random walk and of percolation
clusters. In our case, since the dimension of the islands is $2$, it
suggests that they are compact unlike the clusters in critical
percolation.
\\Statistical behavior of the coastlines here is similar to the
statistics of rocky shorelines studied recently in \cite{Boffetta}.
The winding angle statistics of the shorelines is consistent with
the prediction of SLE$_{8/3}$. To be more serious about the
similarities between the 2D-KPZ coastlines and rocky shorelines and,
moreover, giving another justification for conformal invariancy, we
compute the winding angle statistics for the contour lines of 2D-KPZ
surface.

\begin{figure}\begin{center}
\includegraphics[scale=0.38]{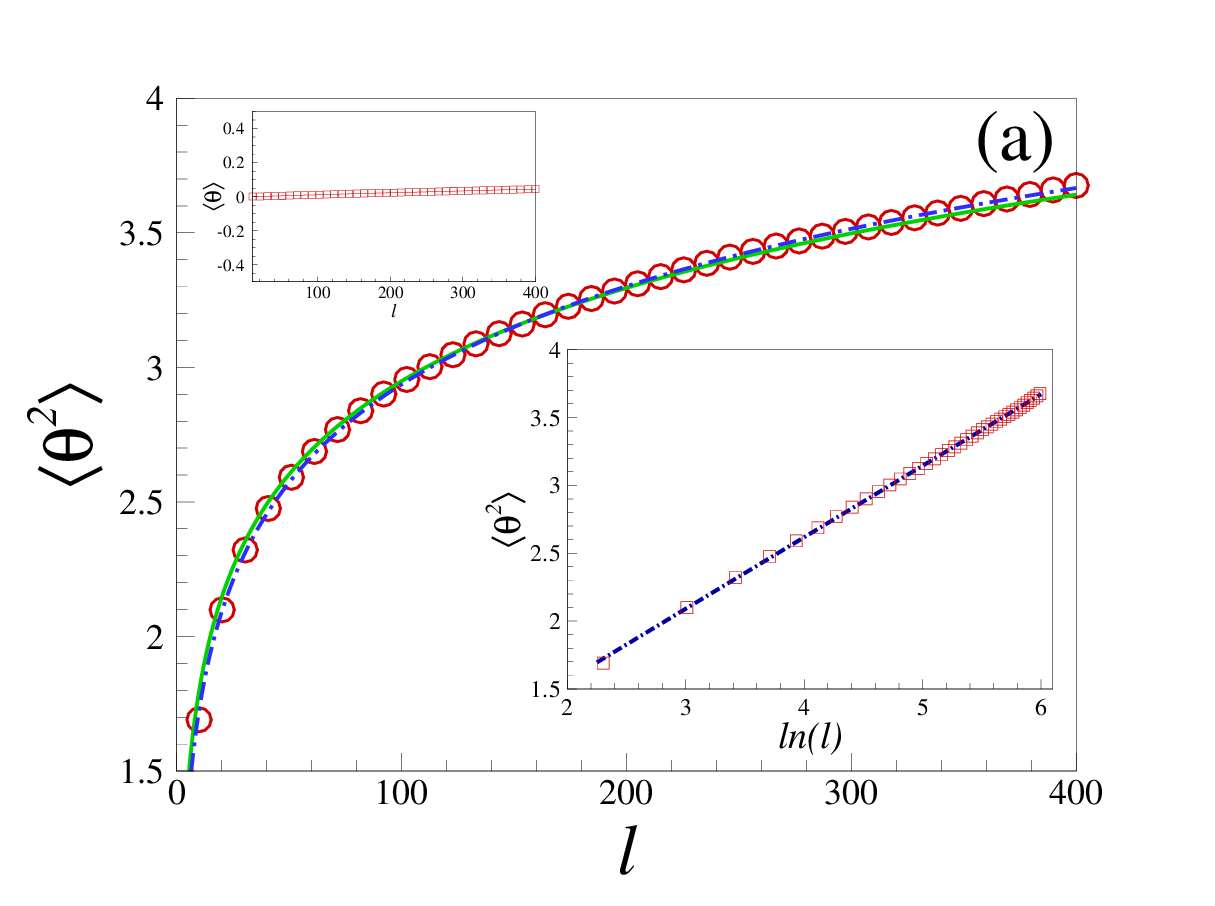}\ \includegraphics[scale=0.38]{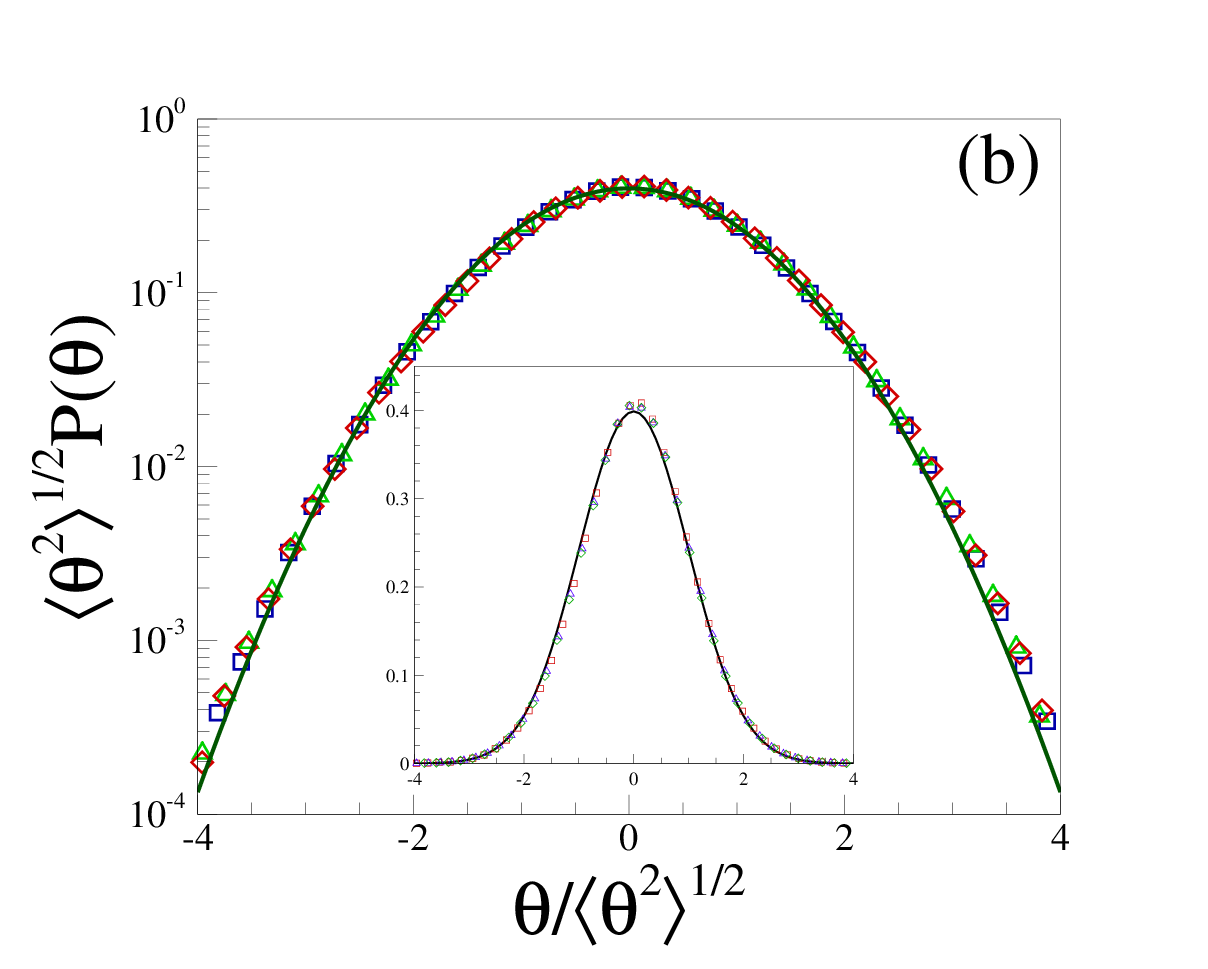}
\narrowtext\caption{\label{Fig5}(color online). Winding angle
statistics of iso-height lines of 2D-KPZ surface simulated on square
lattice of size $2048^2$. (a) Main: logarithmic behavior of the
variance of the winding angle as a function of distance $l$ along
the curve (see Eq. (\ref{winding})). Upper-left inset: the mean
winding angle as a function of $l$. Lower-right inset:
semilogarithmic behavior of the winding. Dotted-dashed lines show
the best fit according to Eq. (\ref{winding}), with
$\kappa=2.76\pm0.1$. The Solid line in the main frame shows the fit
with $\kappa=8/3$ for comparision with SAWs. (b) The rescaled
probability density function of the winding angle at lengths $l$ =
50, 200, 400 in units of lattice spacing compared to the standard
Gaussian density,}\end{center}
\end{figure}

Duplantier and Saleur studied in \cite{Duplantier} the winding angle
between the two endpoints of a finite SAW in two dimensions. Using
Coulomb gas methods, they found that the distribution of winding
angle is Gaussian with the winding variance of $\sim(8/g)\ln L$,
where $L$ is the distance between the endpoints and $g$ is Coulomb
gas coupling parameter which is related to $\kappa$ by $g=4/\kappa$.
They have also shown that the winding angle at a single endpoint
relative to the global average direction of the curve is a Gaussian
with variance of $(4/g)\ln L$. Wieland and Wilson found in
\cite{Wilson} that the variance in the winding at typical points
along the curve is $1/4$ as large as the variance in the winding at
the endpoints.

We define the winding angle $\theta$ (as used in \cite{Boffetta}),
as the angle between the line joining two points separated by a
length $l$ along the curve and the local tangent in the reference
point, measured counterclockwise in $radian$. For conformally
invariant curves of diffusivity $\kappa$, its variance behaves like
\cite{Boffetta} \be\label{winding} \langle\theta^2\rangle \sim
\frac{2\kappa}{8+\kappa}\ln l,\ee where the average is taken over an
ensemble of $1800$ curves by moving the reference point along each
curve.

As shown in Fig. \ref{Fig5}(a), the variance of winding angle for
iso-height lines at a cut made at mean height has a logarithmic
behavior. The best fit (dotted dashed lines), corresponds to
$\kappa=2.76\pm0.1$ which is compared with the fit by setting
$\kappa=8/3$ for SAWs (solid line). The length scale , $l$, is
measured in units of lattice spacing which is set to unity, on
square sample size of $2048^2$. For each configuration the largest
loop is selected, so within the length scale $l$, the curves do not
have a preferred direction and the mean winding angle is zero
(upper-left inset in Fig. \ref{Fig5}(a)). The rescaled probability
density function of the winding angle at different lengths is shown
in Fig. \ref{Fig5}(b), which converges to a standard Gaussian
density. The results are consistent with \cite{saberi}. We find no
changes in the winding statistics of contour lines at different
level sets.\\
In conclusion, studying the statistics of iso-height lines of
saturated 2D-KPZ surface at different level sets we find that the
fractal dimensions of cluster heights and their perimeter remain
unchanged when changing the height of the cut. We also find that the
exponent associated with the distribution function of the cluster
size (the mass of clusters is considered here) changes as a function
of the height of the cut. It linearly decreases for cuts made below
the mean height and crosses over to an almost linear fluctuation
around a specific value above the mean height. We also tested
another exponents related to the distribution of length of the
loops, the area of the loops and the radius of gyration which all
change at different level cuts (the results are not included in this paper).\\
The winding angle statistics of the contour lines also suggests that
their statistics is comparable to that of SAWs. This confirms that
the conformal invariant property of the contour lines are given by
SLE$_{8/3}$.

\end{document}